\documentclass[11pt]{article}

\usepackage[margin=1in]{geometry}
\usepackage[numbers,sort&compress]{natbib}
\bibpunct[, ]{[}{]}{,}{n}{,}{,}

\usepackage{hyperref}
\usepackage{amsmath}
\usepackage{amssymb}
\usepackage{booktabs}
\usepackage{graphicx}

\title{Joint Conditional Maximum Likelihood Estimation of Beta\\
Regression for Panel Data with Higher-Order\\
Autoregressive Errors}

{\small
\author{Ariel Linden, DrPH\\
University of California, San Francisco\\
Department of Medicine\\
Division of Clinical Informatics \& Digital Transformation (DoC-IT)\\
San Francisco, CA, USA\\
ariel.linden@ucsf.edu}
}
\date{}

\begin{document}

\maketitle

\begin{abstract}
Existing panel beta regression estimators do not accommodate
higher-order autoregressive dependence within a joint likelihood
framework. We develop such an estimator: a joint conditional
maximum-likelihood beta-AR($k$) model extended from a single time
series to panel data, and extend it to panel inference through two
new covariance estimators -- a generalization of Beck--Katz
panel-corrected standard errors (PCSE) and a Driscoll--Kraay-style
alternative (DK) -- to account for the cross-panel dependence the
extension introduces. The estimator is evaluated by Monte Carlo
simulation spanning AR(1)--AR(3) error structures, three panel sizes,
and five series lengths, under both independent panels and genuine
cross-panel dependence, against an uncorrected baseline.
PCSE generally exhibited superior Type~I error control and coverage relative to DK, at a
power cost that was modest and partly illusory given DK's own
miscalibration. Under genuine cross-panel dependence, the
uncorrected estimator's Type~I error was severe and did not diminish
with additional data, while both corrections resolved it, with PCSE
again preferred. An illustrative multi-site disease management
example is provided, along with a Stata implementation
(\texttt{xtbetark}).
\end{abstract}


\section{Introduction}

Panel and cross-sectional time-series (CSTS) data --- repeated
observations on a fixed set of units such as hospitals, health
systems, regions, or firms --- are common across the social, health,
and policy sciences \cite{hsiao2003,baltagi2021}. Many of these
applications track outcomes that are proportions or rates:
readmission rates, infection rates, prescribing rates, and similar
population-level measures aggregated within each unit at each time
point. This structure arises naturally in panel extensions of the
interrupted time series (ITS) design, where several units --- several
health systems adopting a policy at the same time, for instance --- are
observed repeatedly before and after an intervention
\cite{linden2015}. Designs like this sit squarely in the moderate-$N$,
substantial-$T$ setting described for panel-corrected inference
\cite{beckkatz1995}. That setting brings about two
separate estimation problems at once --- bounded-outcome regression
and time-series cross-section inference --- and neither can be safely
ignored.

The first problem is the bounded support of the outcome itself.
Fitting an ordinary least squares (OLS) model to a raw proportion is a
form of misspecification in its own right, distinct from ignoring
autocorrelation but no less consequential. A linear model can produce
fitted values outside the admissible $(0,1)$ range, and it disregards
the well-documented relationship between the mean and variance of a
bounded outcome \cite{ferraricribarineto2004}. Beta regression
addresses this directly: it models the conditional mean through a
link function while simultaneously modeling precision, avoiding both
the boundary problem and the heteroskedasticity that comes with it.

The second problem is serial dependence. Time series data collected
at regular intervals are routinely autocorrelated. Ignoring this
structure produces biased standard errors, inflated Type I error, and
unreliable confidence interval coverage, and the severity of the
problem grows with the order and persistence of the underlying
autoregressive process \cite{turner2021,bottomley2023,zhang2011}. Most of this
evidence, however, is limited to AR(1) processes evaluated in
single-group designs. The only studies to date examining higher-order
AR structure directly have done so within multiple-group interrupted
time series designs \cite{lindenmgitsapower2026,lindenmgitsaadjust2026,lindenmgitsajscs2026,lindenxtpraiskpaper2026},
those that naturally comprise a panel data structure, i.e. several groups observed repeatedly over time. This is the only existing body of evidence on how serial-dependence corrections perform beyond the first-order case, and it motivates
this paper's own emphasis on AR orders one through three rather than
AR(1) alone. A closed-form recursive substitution accommodating
autoregressive dependence of arbitrary order within the beta
regression likelihood itself, rather than as a post-hoc correction,
was later addressed directly \cite{rochacribarineto2009,ferreira2015}. The community
contributed program \texttt{betark} \cite{lindenbetark} implements
the resulting recursive conditional likelihood in Stata, jointly
estimating the mean equation, the precision equation, and the AR($k$)
coefficient vector by maximum likelihood. Its Monte Carlo evaluation
showed that this joint approach out-calibrates the standard
alternative --- a quasi-binomial GLM with Newey--West HAC standard
errors --- and that the advantage grows under more persistent
autocorrelation \cite{lindenbetarkpaper2026}.

Neither strand, on its own, addresses the panel setting. When
observations are nested within units --- the case motivating this
paper --- a third source of dependence enters. Errors may be
heteroskedastic across units, and units subject to common shocks or
shared environments may be correlated with one another at a given
point in time \cite{beckkatz1995}. \texttt{betark}, like the
beta-AR($k$) literature more broadly, was built for a single time
series. It has no mechanism for this cross-sectional structure. No
existing estimator addresses bounded support, within-unit serial
dependence, and between-unit dependence all at once. This paper
develops a joint conditional maximum likelihood estimator for beta
regression with panel-specific higher-order autoregressive errors.

This gap has already been closed for continuous outcomes. The
single-series
Prais--Winsten AR($k$) estimator was extended to panel data by pairing Prais--Winsten's exact GLS transformation with the panel-corrected standard error (PCSE) sandwich \cite{beckkatz1995} and implemented in Stata via the community contributed \texttt{xtpraisk} package \cite{lindenxtpraiskpaper2026,linden2026xtpraisk}. This parametric approach was then compared against the natural nonparametric alternative for panel time series --- Driscoll and Kraay's
HAC-based estimator (DK) \cite{driscollkraay1998}. Across a wide range of autoregressive orders,
panel sizes, and series lengths, the parametric PCSE-based approach
achieved higher power without sacrificing inferential validity. That
advantage widened under more persistent autocorrelation and higher AR
order --- exactly the conditions where a nonparametric HAC correction
has the least information to work with.

This paper extends \texttt{betark} to panel data in the same spirit as 
the panel Prais--Winsten extension above, with the new community contributed 
\texttt{xtbetark} package for Stata \cite{lindenxtbetark}. One fundamental difference distinguishes this
single-series-to-panel-data extension from that of the Prais---Winsten case.
Beck and Katz's PCSE sandwich was built for a linear GLS estimator, where the algebra separates cleanly into a scalar residual and a design matrix. \texttt{betark} is a nonlinear joint maximum likelihood estimator with three sets of parameters --- mean, precision, and AR coefficients --- and nothing
that plays the role of a residual in the PCSE sense. Generalizing
PCSE to this setting means reconstructing the sandwich from
equation-level likelihood scores instead, applied separately to every
pair of equations rather than once. Whether this generalization, and
its Driscoll--Kraay-style counterpart, preserve the same ranking
found for continuous outcomes --- where the parametric correction was the clear preference \cite{lindenxtpraiskpaper2026} --- is not something that earlier result guarantees.
It is the empirical question this paper investigates.

This paper makes three contributions. First, we extend
\texttt{betark}'s joint conditional beta-AR($k$) likelihood from a
single time series to panel data. Second, because pooling the
likelihood correctly across panels is not, on its own, sufficient for
valid inference, we develop two distinct approaches for handling the
resulting cross-panel dependence: a generalization of the Beck--Katz
PCSE sandwich to a multi-equation joint likelihood, and a
Driscoll--Kraay-style nonparametric alternative. Third, we test both
approaches in a comprehensive Monte Carlo simulation spanning
autoregressive orders one through three, three autocorrelation
scenarios per order, a range of panel sizes and series lengths, and
multiple effect sizes, under both panel-independent and genuinely
cross-panel-dependent data-generating processes. This evaluation asks
a question neither of the two studies motivating this paper could
answer on its own. Does the performance advantage PCSE
demonstrated for continuous outcomes \cite{lindenxtpraiskpaper2026},
and the calibration advantage the joint likelihood demonstrated for a
single time series \cite{lindenbetarkpaper2026}, survive once
the two extensions --- panel data and a bounded, nonlinear outcome
model --- are combined? We further provide an illustrative example
showing what estimator choice means in practice, in a realistic
panel-ITS setting.

The rest of the paper proceeds as follows. Section~2 describes the
beta-AR($k$) panel model, the two variance-covariance estimators, and
the simulation strategy used to evaluate them. Section~3 presents the
simulation results. Section~4 gives an illustrative example.
Section~5 discusses the findings, and Section~6 concludes.

\section{Methods}

\subsection{The Beta-AR($k$) Panel Model}

\subsubsection{Model}

Let $y_{it} \in (0,1)$ denote the outcome for panel $i = 1,\dots,N$
at time $t = 1,\dots,T_i$. We model $y_{it}$ as beta-distributed conditional on a mean
parameter $\mu_{it} \in (0,1)$ and a precision parameter
$\phi_{it} > 0$, following the standard beta regression
parameterization \cite{ferraricribarineto2004},
\[
y_{it} \mid \mu_{it}, \phi_{it} \sim \text{Beta}\big(\mu_{it}\phi_{it},\,(1-\mu_{it})\phi_{it}\big),
\]
with density
\[
f(y_{it}) = \frac{\Gamma(\phi_{it})}{\Gamma(\mu_{it}\phi_{it})\,\Gamma((1-\mu_{it})\phi_{it})}\,
y_{it}^{\mu_{it}\phi_{it}-1}(1-y_{it})^{(1-\mu_{it})\phi_{it}-1}.
\]
Under this parameterization $E(y_{it}) = \mu_{it}$ and
$\mathrm{Var}(y_{it}) = \mu_{it}(1-\mu_{it})/(1+\phi_{it})$, so $\phi_{it}$
governs precision directly: larger values imply a tighter conditional
distribution around $\mu_{it}$ for a given mean. Both parameters are
modeled through their own linear predictor and link function,
\[
g(\mu_{it}) = x_{it}'\beta, \qquad h(\phi_{it}) = z_{it}'\gamma,
\]
with $g(\cdot)$ the logit link and $h(\cdot)$ the log link, matching
the defaults of Stata's \texttt{betareg} and \texttt{betark}.

Serial dependence enters through the mean equation only, via a
recursive substitution accommodating AR($k$) dependence within the
likelihood itself \cite{rochacribarineto2009,ferreira2015}. Writing $\eta_{it} = x_{it}'\beta$
for the deterministic linear predictor, the AR($k$) recursion defines
an adjusted linear predictor
\begin{equation}
\eta_{it}^{*} = \eta_{it} + \sum_{j=1}^{k} \rho_j \big[g(y_{i,t-j}) - \eta_{i,t-j}\big],
\label{eq:recursion}
\end{equation}
and the conditional mean actually used to parameterize the beta
density at $(i,t)$ is $\mu_{it}^{*} = g^{-1}(\eta_{it}^{*})$. The
bracketed term in \eqref{eq:recursion} is the beta-scale analogue of
an autoregressive error: the difference between the observed,
logit-transformed lag and its own deterministic prediction. Because
this substitution is made directly in the link-scale linear
predictor rather than as a correction applied after estimation, the
resulting joint conditional likelihood
\begin{equation}
\ell(\beta,\gamma,\rho) = \sum_{i=1}^{N} \sum_{t=k+1}^{T_i} \log f\big(y_{it}\,;\,\mu_{it}^{*},\phi_{it}\big)
\label{eq:jointlik}
\end{equation}
estimates the mean equation, precision equation, and AR($k$)
coefficient vector $\rho = (\rho_1,\dots,\rho_k)'$ jointly, by maximum
likelihood, rather than treating autocorrelation as a nuisance to be
corrected for afterward.

The first $k$ observations of every panel carry no AR adjustment (a
cold start, since $g(y_{i,t-j})-\eta_{i,t-j}$ is undefined before
$t>k$) and do not contribute to \eqref{eq:jointlik}. \texttt{betark}
restarts this cold start at every gap in the time index within a
series, so that a single series with missing periods is treated as
several independent AR($k$) segments rather than one contaminated by
a spurious lag across the gap. \texttt{xtbetark} requires no change to
this mechanism: a panel change is, from the recursion's point of
view, simply another gap. Declaring the data with both a panel and a
time variable and restarting the cold start at every panel-id change
in addition to every time gap is sufficient to pool
\eqref{eq:jointlik} correctly across panels. The point-estimation
machinery of \texttt{betark} and \texttt{xtbetark} is therefore
identical; the two commands differ only in what follows.

\subsubsection{Estimation}

Both \texttt{betark} and \texttt{xtbetark} commands maximize \eqref{eq:jointlik} by Newton--Raphson, with a BHHH step as a fallback when the Hessian is not well behaved, using Stata's \texttt{moptimize()}. Starting values are taken from a static beta regression fit ($\rho = 0$) on the same mean and scale specification. Let $\theta = (\beta',\gamma',\rho')'$ denote the full parameter vector, of dimension $q = k_1+k_2+k$ where $k_1$ and $k_2$
are the number of mean- and scale-equation coefficients (the score
construction below is indexed instead over $m=2+k$ linear indices,
one per equation; see Section~\ref{sec:chainrule}). The default
covariance is the observed-information matrix, $\widehat{V}_{\text{OIM}} = H^{-1}$, where $H$ is the negative Hessian of \eqref{eq:jointlik} at
convergence. This is the covariance \texttt{betark} reports, and it
implicitly assumes panels are independent of one another --- an
assumption a genuine panel structure need not satisfy.

\subsubsection{A Chain-Rule Score for a Multi-Equation Likelihood}
\label{sec:chainrule}

Both panel-corrected variance estimators below are built from the
per-observation score of \eqref{eq:jointlik}, and both require a
detail specific to how Stata's \texttt{moptimize()} evaluates a
likelihood specified, as \eqref{eq:jointlik} is, through a set of
linear indices rather than through the coefficient vector directly.
For an evaluator of this type, \texttt{moptimize\_result\_scores()}
returns the score with respect to each of the $m = 2+k$ linear
indices --- $\eta_{it}$, $\log\phi_{it}$, and each $\rho_j$ (which,
having no covariates of its own, is already its own linear index) ---
not the score with respect to each of the $q$ coefficients. Obtaining
a per-coefficient score requires an explicit chain-rule step,
\begin{equation}
g_{it}^{(e)} = s_{it}^{(e)} \cdot x_{it}^{(e)},
\label{eq:chainrule}
\end{equation}
where $s_{it}^{(e)} = \partial \ell_{it}/\partial(\text{linear index }e)$
is the equation-level score and $x_{it}^{(e)}$ is that equation's own
design row (the mean equation's covariates for $e=1$, the scale
equation's covariates for $e=2$, and simply $1$ for each AR equation,
since \eqref{eq:chainrule} is then just $s_{it}^{(e)}$ itself).
Stacking $g_{it}^{(1)},\dots,g_{it}^{(m)}$ gives the full
$q$-dimensional per-coefficient score $g_{it}$ for observation $(i,t)$.

\subsubsection{Panel-Corrected Standard Errors}

Beck and Katz's PCSE sandwich \cite{beckkatz1995}, as
implemented for a single linear equation in \texttt{xtpraisk}
\cite{lindenxtpraiskpaper2026}, estimates a contemporaneous
cross-panel covariance from a single scalar residual series and
combines it with the model's design matrix. \texttt{xtbetark} has
three equations rather than one, and no residual in the PCSE sense --
only the equation-level scores of Section~\ref{sec:chainrule}.
Generalizing PCSE to this setting means building an analogous
covariance and design cross-product for every pair of equations, not
just once. For equations $e,e' \in \{1,\dots,m\}$ and panels $i,j$,
define
\begin{equation}
\widehat{\Sigma}^{(e,e')}_{ij} = \frac{1}{T_{ij}} \sum_{t=1}^{T_{ij}} s_{it}^{(e)}\, s_{jt}^{(e')},
\label{eq:pcsesigma}
\end{equation}
where $T_{ij} = \min(T_i,T_j)$, matching the first $T_{ij}$ time
periods of each panel, exactly as in the single-equation case.
Following standard sandwich-estimator terminology --- $H^{-1}$,
appearing on both outer sides of the final expression below, is the
sandwich's \emph{bread}; the matrix sitting between the two bread
terms is its \emph{meat} --- the corresponding $(e,e')$ block of the
$q \times q$ meat matrix is
\begin{equation}
M^{(e,e')} = \sum_{i=1}^{N}\sum_{j=1}^{N} \widehat{\Sigma}^{(e,e')}_{ij}\, X_i^{(e)\prime} X_j^{(e')},
\label{eq:pcsemeat}
\end{equation}
where $X_i^{(e)}$ is panel $i$'s design matrix for equation $e$. Since
swapping $(e,i) \leftrightarrow (e',j)$ in \eqref{eq:pcsesigma}--\eqref{eq:pcsemeat}
shows $M^{(e',e)} = (M^{(e,e')})'$, only the $\binom{m}{2}+m$ blocks
with $e \le e'$ need to be computed directly; the remainder follow by
transposition. Assembling all blocks gives the full meat matrix $M$,
and
\begin{equation}
\widehat{V}_{\text{PCSE}} = H^{-1} M H^{-1}.
\label{eq:pcsevce}
\end{equation}
As in the original Beck--Katz construction, \eqref{eq:pcsesigma} is
purely contemporaneous: no lag of $t$ appears anywhere in the
covariance. This is a deliberate match to a model whose AR($k$)
component already removes serial dependence parametrically, a point
we return to in Section~\ref{sec:dk}.

\paragraph{Why scores, not residuals.} The choice of the equation-level
score $s_{it}^{(e)}$ as the object entering
\eqref{eq:pcsesigma}--\eqref{eq:pcsemeat}, rather than a residual,
Pearson residual, or influence function, follows directly from
general M-estimation sandwich theory \cite{huber1967,white1982}
rather than from analogy alone. For any M-estimator defined by
solving $\sum_i s_i(\theta) = 0$ for a score function $s_i$, the
asymptotic variance of $\hat\theta$ takes the sandwich form $H^{-1}
\left[\lim_{N\to\infty} N^{-1}\mathrm{Var}\!\left(\sum_i
s_i(\theta_0)\right)\right] H^{-1}$ regardless of whether the
estimating equations arise from a linear model, a generalized linear
model, or, as here, a multi-equation joint likelihood. Ordinary least
squares is the special case in which this general theory happens to
coincide with a residual-based formula: OLS's score contribution is
$x_{it}\hat u_{it}$, so the residual is the score with the design row
factored out, and Beck and Katz's
residual-based $\widehat\Sigma_{ij}$ combined with $X_i'X_j$ \cite{beckkatz1995} is
algebraically identical to the general sandwich meat evaluated at the
OLS score. \eqref{eq:pcsesigma}--\eqref{eq:pcsemeat} is therefore not
an analogy to the Beck--Katz construction but the same general
sandwich meat evaluated at this model's own score, which is the
equation-level $s_{it}^{(e)}$ rather than a residual because the
model has no single residual to factor a design row out of. Pearson
residuals, generalized residuals, and other model-diagnostic
quantities do not satisfy the defining M-estimator property that
their panel sums have expectation zero at $\theta_0$; substituting
any of them in place of $s_{it}^{(e)}$ would not yield a sandwich
estimator with the stated asymptotic guarantee. Influence functions
are asymptotically equivalent to $H^{-1}s_i$ rather than a competing
choice: sandwiching $H^{-1}$ around $\mathrm{Var}(\sum_i s_i)$, as
\eqref{eq:pcsevce} does, and averaging pre-multiplied influence
functions directly are the same estimator expressed two ways.

\paragraph{Consistency.} Two conditions make \eqref{eq:pcsemeat} a
consistent estimator of the true sandwich meat. First, pooling
independent panels means the joint score is the sum of per-panel
scores, so standard MLE asymptotic normality applies to $\hat\theta$
as $N \to \infty$ under the usual M-estimator regularity conditions --
identifiability of $\theta_0$, twice-continuous differentiability of
the log-likelihood, and finite fourth moments of the per-panel score
sum -- exactly as it would for any pooled independent-cluster
M-estimator; nothing about the AR($k$) structure or the beta
conditional density changes this step. Second, and specific to
\eqref{eq:pcsesigma}, Beck and Katz's own framework is explicitly a
large-$T$, moderate-$N$ asymptotic theory: $\widehat\Sigma^{(e,e')}_{ij}$
is a time-average that converges to the true cross-panel score
covariance as $T \to \infty$ under standard mixing conditions on the
within-panel score process. Those conditions follow from the AR($k$)
process being stationary, which the simulation design enforces
directly by excluding replications whose generated AR coefficients
would imply a non-stationary process (Section~\ref{sec:simdesign}).
\eqref{eq:pcsesigma}--\eqref{eq:pcsemeat} therefore inherits exactly
the asymptotic justification Beck and Katz's own estimator relies on,
applied to this model's score rather than to an OLS residual.

\paragraph{Scope.} These results assume the beta-AR($k$) likelihood
is correctly specified for the conditional mean and precision
structure. Like any sandwich estimator, PCSE and DK protect against
misspecification of the \emph{dependence} structure -- contemporaneous
cross-panel correlation for PCSE, serial and cross-panel correlation
for DK -- given a correctly specified mean model; neither protects
against a wrong link function, an omitted covariate, or a
misspecified conditional distribution, and this study does not
evaluate that separate class of misspecification. Two further
technical points, neither of which arose in the conditions examined
here, are worth stating explicitly rather than leaving implicit.
Numerically, the meat matrix $M$ could in principle become
ill-conditioned or rank-deficient when $T$ is small relative to the
number of estimated coefficients, particularly at higher AR orders;
this risk is inherent to the Beck--Katz construction generally, not
specific to this extension, and was not encountered in any
replication reported here ($T \geq 30$ throughout), but nor was it
specifically stress-tested at more extreme combinations. Separately,
this study's simulations focused on inferential properties --
Type~I error, coverage, and power -- rather than computational
performance, so optimizer convergence failures were not
systematically summarized across the full simulation grid beyond the
stationarity-based exclusion already applied to the data-generating
process. Whether Newton--Raphson or BHHH failed to converge for a
nontrivial share of replications at the most difficult design cells
(short $T$, high AR order, high persistence) is accordingly not
reported here, and remains an open question rather than a claim that
no such failures occurred.

Thus, the proposed PCSE estimator is not an analogy to Beck and
Katz's residual-based estimator \cite{beckkatz1995} but its
natural M-estimation generalization to a multi-equation likelihood.

\subsubsection{A Driscoll--Kraay-Style Alternative}
\label{sec:dk}

\texttt{xtbetark} also implements a
Driscoll--Kraay-style \cite{driscollkraay1998} sandwich, following
a construction proposed for panel-corrected inference generally
\cite{hoechle2007} but applied to the
chain-ruled score $g_{it}$ of Section~\ref{sec:chainrule} in place of
the ordinary least squares score it was originally built for. Let
\begin{equation}
h_t = \sum_{i:\,t \in i} g_{it}
\label{eq:dkht}
\end{equation}
be the cross-sectional sum of the per-coefficient score over all
panels present at time $t$. The sandwich meat is
\begin{equation}
\Omega = \sum_{t} h_t h_t' + \sum_{j=1}^{L} w_j \Big(\sum_{t>j} h_t h_{t-j}' + \big(\sum_{t>j} h_t h_{t-j}'\big)'\Big),
\qquad w_j = 1 - \frac{j}{L+1},
\label{eq:dkomega}
\end{equation}
a Bartlett kernel over time lags up to bandwidth $L$, and
$\widehat{V}_{\text{DK}} = H^{-1}\Omega H^{-1}$. Unlike PCSE,
\eqref{eq:dkomega} smooths over serial as well as cross-panel
dependence. We set the default bandwidth to $L=k$, the fitted AR
order, rather than the Newey--West automatic
rule $\lfloor 4(T/100)^{2/9}\rfloor$ \cite{newey1994} that this construction uses by
default in most panel-regression applications. That rule is calibrated for ordinary least squares residuals
with no parametric correction for serial dependence at all; for a
model whose AR($k$) likelihood has already removed serial dependence
from the score, whatever residual correlation remains should
plausibly not extend meaningfully beyond the fitted lag order, and
should not grow with $T$ the way it would in the absence of any
parametric correction. Because \eqref{eq:dkomega} smooths over a
structure the likelihood has, under correct specification, already
addressed, we expect $\widehat{V}_{\text{DK}}$ to be more prone than
$\widehat{V}_{\text{PCSE}}$ to the finite-sample anticonservative bias
well documented for Newey--West-type HAC estimators when little
genuine serial correlation remains for the kernel to explain --- the
central hypothesis this paper's simulation study is designed to test.
This choice of $L=k$ was fixed a priori on the reasoning above rather
than selected empirically; we did not compare it against $L=k+1$,
$L=2k$, or the automatic Newey--West rule within this study, and a
direct bandwidth-sensitivity comparison remains an open question this
paper does not resolve.

Both variance estimators share the same bread, $H^{-1}$, computed once
per model fit; \texttt{xtbetark} obtains $\widehat{V}_{\text{OIM}}$,
$\widehat{V}_{\text{PCSE}}$, and $\widehat{V}_{\text{DK}}$ from a
single optimization rather than one fit per variance estimator, since
only the sandwich construction on top of an already-completed fit
differs between them.

\subsection{Simulation Strategy}
\label{sec:simstrategy}

We evaluate \texttt{xtbetark}'s two panel-corrected variance
estimators against the uncorrected $\widehat{V}_{\text{OIM}}$ ---
numerically identical to fitting \texttt{betark} directly on the same
panel data, since the two commands share the same
point-estimation machinery --- across two simulation designs,
following the same overall structure used to evaluate
\texttt{betark} \cite{lindenbetarkpaper2026} and \texttt{xtpraisk}
\cite{lindenxtpraiskpaper2026}: a primary design characterizing
baseline performance, and a sensitivity design stress-testing the two
corrections under genuine cross-panel dependence, the condition
neither \texttt{betark} nor a primary design with independent panels
can speak to.

\subsubsection{Data-Generating Process}

Panels are generated from the same beta-AR($k$) recursion described
in \eqref{eq:recursion}. The deterministic component of the
mean-equation linear predictor, $\eta_{it} = x_{it}'\beta$, follows
the standard regression-based single-group interrupted time series (ITS)
specification \cite{linden2015},
\begin{equation}
\eta_{it} = \beta_0 + \beta_1 t + \beta_2 D_{it} + \beta_3 D_{it}(t-t_0),
\label{eq:itsa}
\end{equation}
where $D_{it} = \mathbb{1}(t \ge t_0)$ indicates the post-intervention
period and $t_0$ is the intervention time point, fixed at the series
midpoint in every design cell. In this parameterization, $\beta_0$ is
the baseline level at $t=0$, $\beta_1$ is the pre-intervention trend,
$\beta_2$ is the immediate level change at the intervention, and
$\beta_3$ is the change in slope following the intervention --- the
difference between the post- and pre-intervention trends. It is
$\beta_3$ that is the treatment effect of primary inferential interest
throughout this study: the coefficient on which Type~I error, power,
$95\%$ coverage, and bias are all evaluated in Section~3.

Rather than specifying $\beta_0,\dots,\beta_3$ directly, we
parameterize \eqref{eq:itsa}, as in \texttt{betark}'s own simulation
study, by percentage change on the response scale: a starting mean
$\mu_0 = g^{-1}(\beta_0)$, a pre-intervention trend, an immediate
post-intervention level change, and a post-intervention trend change,
each expressed as a percentage of $\mu_0$ and converted to their
corresponding logit-scale $\beta$ coefficients before generating each
panel. All panels share an identical deterministic mean structure and
intervention timing; only the AR($k$) innovations differ across
panels under the primary design. We fix $\mu_0 = .10$, a coefficient
of variation of $.05$, no pre-intervention trend, and $\beta_2 = 0$
(no immediate level change) throughout, so that $\beta_3$ is isolated
as the only non-null component of \eqref{eq:itsa} manipulated across
effect-size conditions.

For the sensitivity design, genuine cross-panel dependence is
introduced by injecting a shared common factor into the deterministic
linear predictor before the AR($k$) recursion is applied,
\[
\eta_{it} = x_{it}'\beta + \lambda f_t, \qquad f_t \overset{\text{iid}}{\sim} N(0,1),
\]
with $f_t$ drawn once per time period and shared across all $N$
panels at that period --- the structural analogue, for a model that
is not additive on the outcome scale, of the additive common-factor
design $x_{it} = \lambda f_t + v_{it}$ used to stress-test
\texttt{xtpraisk} \cite{lindenxtpraiskpaper2026}. We use loadings
$\lambda \in \{0.5, 1.0\}$.

\subsubsection{Design}
\label{sec:simdesign}

We cross three autoregressive orders ($k=1,2,3$) with three
autocorrelation scenarios per order --- mild positive, oscillatory,
and high persistent --- panel sizes $N \in \{10,15,20\}$, series
lengths $T \in \{30,60,100,200,400\}$, and post-intervention effect
sizes of $\{0,2,4,8\}\%$ change in the mean, with $0\%$ corresponding
to the null. This panel-size range reflects the moderate-$N$,
substantial-$T$ regime Beck and Katz's own
asymptotic framework targets \cite{beckkatz1995}, rather than the large-$N$,
short-$T$ micro-panels common in economics, and matches the scale of
the applied setting motivating this paper: multi-site health-system
or policy interventions with a limited number of participating
practices or jurisdictions observed over many periods
(Section~\ref{sec:example} illustrates one such example with
$N=12$). $N=5$ was judged too small to meaningfully estimate the
$N \times N$ contemporaneous covariance structure PCSE requires; $N
\geq 50$ would move outside Beck and Katz's own intended regime and
into settings where alternative panel-corrected estimators designed
for large-$N$ panels may be more appropriate. Crossing these factors with the three variance
estimators under comparison ($\widehat{V}_{\text{OIM}}$,
$\widehat{V}_{\text{PCSE}}$, $\widehat{V}_{\text{DK}}$) yields
$3\times3\times5\times4\times3 = 540$ evaluated design cells per
autoregressive order, or $1{,}620$ in total across all three orders,
each summarized from $2{,}000$ replications. The sensitivity design
uses the same grid restricted to the mild and high-persistent
scenarios only, following
\cite{lindenxtpraiskpaper2026}, but adds the common-factor loading
$\lambda$ as a further dimension. This yields
$2\times3\times5\times2\times4\times3 = 720$ evaluated design cells
per autoregressive order, or $2{,}160$ in total across all three
orders, each likewise summarized from $2{,}000$ replications. Autoregressive coefficients for the high-persistent scenario at orders two and three were chosen to keep the implied autoregressive process a comparable distance from the stationarity boundary as the order-one scenario, rather than reusing $\rho$ values calibrated for a single, unpooled series.

\subsubsection{Performance Measures}

We report, following standard Monte Carlo simulation practice
\cite{burton2006}, for the treatment-effect
coefficient $\beta_3$ and for each estimated autoregressive
coefficient $\rho_j$: the empirical Type~I error rate (rejection rate
under the null of no effect), power (rejection rate under each
non-null effect size), $95\%$ confidence interval coverage,
percentage bias, root mean squared error, and the mean model-based
standard error relative to $\widehat{V}_{\text{OIM}}$'s in the same
replication. All tests use a Wald statistic at $\alpha =.05$. All
analyses were conducted using Stata version 19 \cite{statacorp2025}.

\section{Results}
\label{sec:results}

This section presents Monte Carlo evidence on the calibration of the
two panel-appropriate variance estimators described in
Section~\ref{sec:dk} above -- PCSE and DK -- relative to the
uncorrected (naive, $\widehat{V}_{\text{OIM}}$) baseline. Power,
coverage, Type~I error, and SE ratio are presented graphically.
Power and coverage are shown under AR(2) error structures at $N=10$
in the main text, with the corresponding AR(1) and AR(3) figures
provided in the Supplement; Type~I error and SE ratio are shown
across all three autoregressive orders together, in a single figure
each, since both depend materially on AR order. The panel-size
analysis (Section~\ref{sec:res-panelsize}) is likewise shown under
AR(2) only. Percentage bias and RMSE, both properties of the point
estimate alone (see below), are reported numerically rather than
graphically, and are not repeated in the Supplement. The sensitivity
analysis (Section~\ref{sec:res-sensitivity}) evaluates genuine
cross-panel dependence and is reported separately from the
independent-panel results above.

Both variance estimators under comparison, PCSE and DK, share
identical point estimates within a replication -- they differ only in
how each estimator's standard error is constructed, not in the
underlying coefficient. Bias and RMSE are therefore identical between
PCSE and DK by construction; what differs between methods throughout
this section is calibration (Type~I error, coverage, SE ratio), not
the point estimate itself. The naive estimator is shown for reference
in every figure -- as a flat line at the nominal target for measures
with a fixed target (coverage, Type~I error), and as its own averaged
trajectory (grey, dashed, no markers) for power, the one graphed
measure without a fixed target.

Across every measure reported below, performance moved toward its
expected or nominal value as $T$ increased, but not smoothly at every
step: individual $T$ values sometimes departed from the surrounding
trend rather than falling on a monotonic curve. This pattern is
noted where it appears below and is not otherwise interpreted in this
section.

\subsection{Power}
\label{sec:res-power}

\begin{figure}[htbp]
{\centering \includegraphics[height=0.8\textheight, width=\textwidth, keepaspectratio]{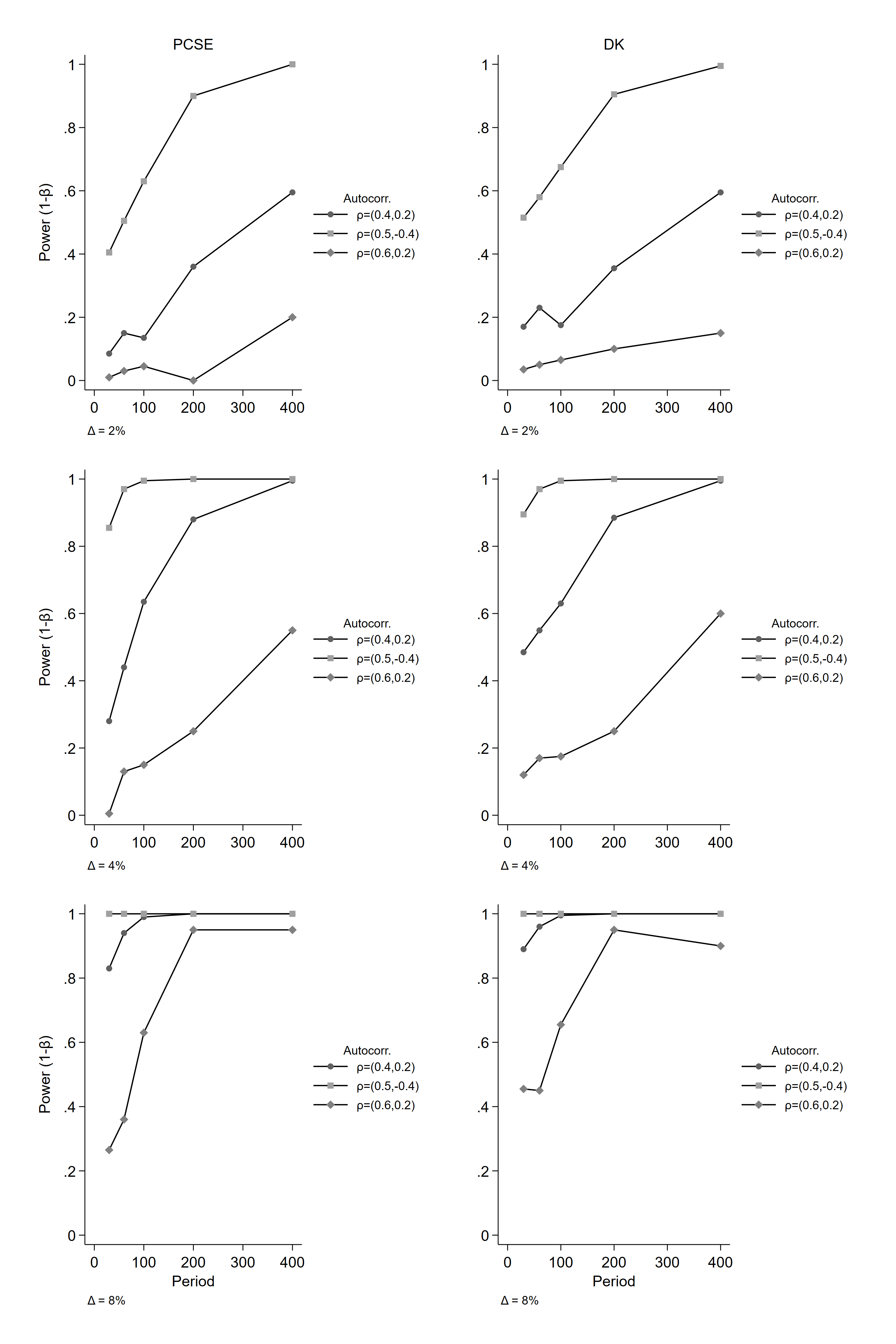}\par}
\caption{Statistical power ($1-\beta$) for PCSE (left column) and DK
(right column) under AR(2) error structures. Rows represent effect
sizes ($\Delta = 2\%, 4\%, 8\%$). Lines distinguish autocorrelation
scenarios: mild positive $\rho=0.4$ (circles); oscillatory
$\rho=-0.4$ (squares); high persistent $\rho=0.6$ (diamonds). $N=10$.}
\label{fig:1}
\end{figure}

Figure~\ref{fig:1} presents statistical power for PCSE and DK under
AR(2) error structures across three effect sizes ($\Delta = 2\%, 4\%,
8\%$) and three autocorrelation scenarios, at $N=10$. Power increased
with $T$ and with effect size for both estimators, though not
monotonically at every step -- individual cells occasionally departed
from the surrounding trend rather than tracing a perfectly smooth
curve. Power was consistently strongest under oscillatory
autocorrelation, reaching $100\%$ by $T=60$ at $\Delta=4\%$ for both
methods, and weakest under high persistent autocorrelation, where
power remained below $50\%$ at $T=30$ even at the largest effect size
($\Delta=8\%$: DK $46\%$, PCSE $26\%$) and did not exceed $95\%$ until
$T=200$. DK showed modestly higher power than PCSE at short $T$
across every scenario -- for example, at $T=30$, $\Delta=2\%$, mild
positive autocorrelation, DK reached $17\%$ power against PCSE's
$8\%$. Naive power, not shown graphically (see note on
Figure~\ref{fig:1}), was $4\%$ at $T=30$, $\Delta=2\%$, mild positive
autocorrelation, and $18\%$ at $T=30$, $\Delta=8\%$, high persistent
autocorrelation, versus DK's $17\%$ and $46\%$ at the same cells. The
corresponding AR(1) and AR(3) results (Supplement, Figures A1--A2)
followed the same qualitative pattern with respect to scenario and
effect size. Power was consistently highest under AR(1) for a given
$T$ and effect size; AR(2) and AR(3) were closer to each other than
either was to AR(1), without a consistent ordering between them
across every $T$ examined.

\subsection{95\% Confidence Interval Coverage}
\label{sec:res-coverage}

\begin{figure}[htbp]
{\centering \includegraphics[height=0.8\textheight, width=\textwidth, keepaspectratio]{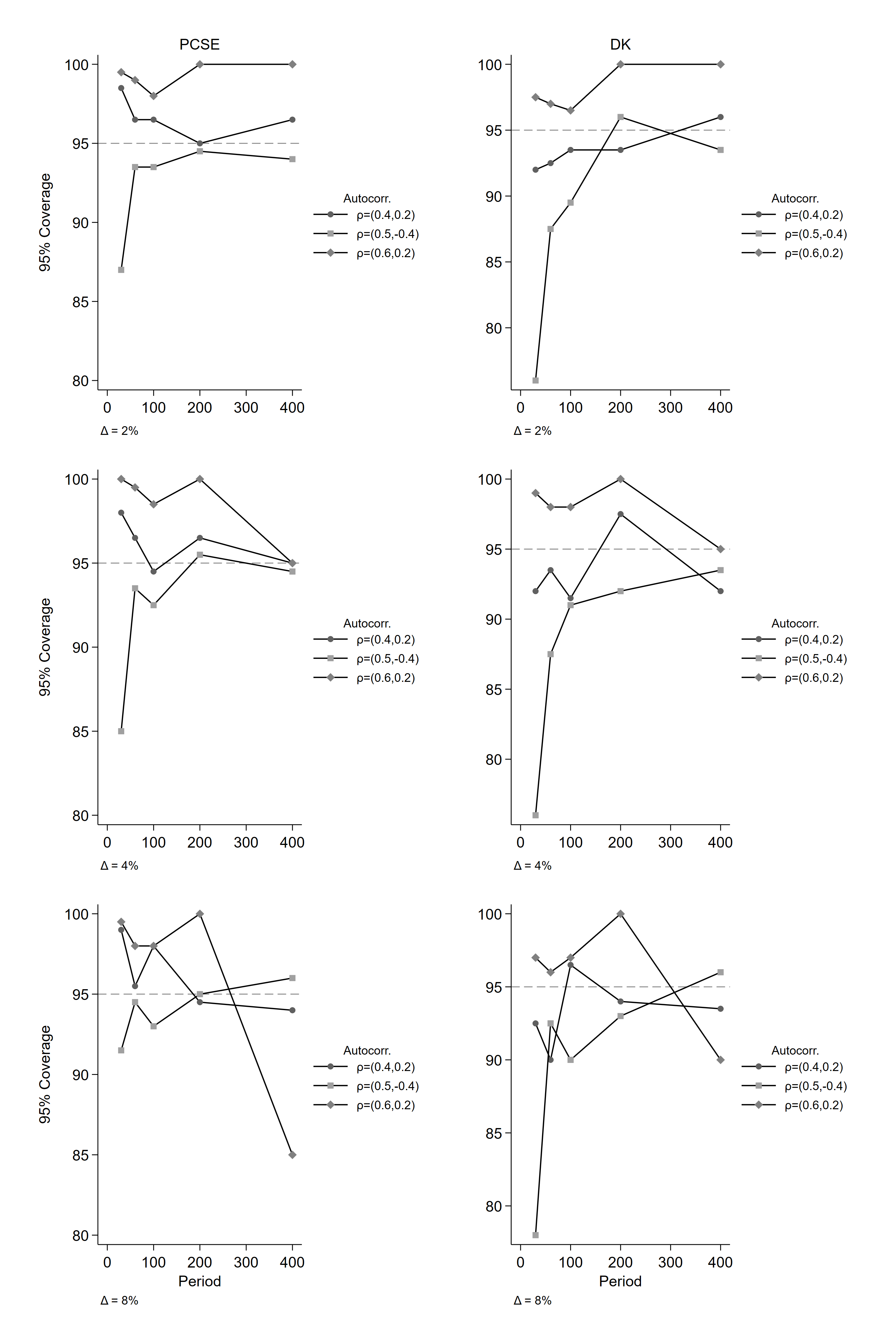}\par}
\caption{$95\%$ confidence interval coverage for PCSE (left column)
and DK (right column) under AR(2) error structures. Rows represent
effect sizes ($\Delta = 2\%, 4\%, 8\%$). Lines distinguish
autocorrelation scenarios: mild positive $\rho=0.4$ (circles);
oscillatory $\rho=-0.4$ (squares); high persistent $\rho=0.6$
(diamonds). Dashed reference line at nominal $95\%$. $N=10$.}
\label{fig:2}
\end{figure}

Figure~\ref{fig:2} presents $95\%$ confidence interval coverage under
AR(2) error structures. Coverage for both PCSE and DK remained
broadly close to the nominal $95\%$ target under mild positive and
high persistent autocorrelation across the full range of $T$, with
cell-to-cell variability rather than a smooth trend: individual cells
ranged as low as $90\%$ and as high as $100\%$ with no consistent
drift in either direction as $T$ increased. The clearest departure
from nominal coverage occurred under oscillatory autocorrelation at
short $T$, where DK's coverage fell to $76\%$ at $T=30$ across all
three effect sizes, compared with PCSE's $85$--$91\%$ over the same
cells. Both estimators' coverage in the oscillatory scenario returned
to within a few points of nominal by $T \geq 100$. Naive coverage
(not shown graphically) tracked close to nominal throughout. AR(1)
and AR(3) results (Supplement, Figures A3--A4) showed the same
oscillatory-short-$T$ pattern for DK, though the departure from
nominal at $T=30$ was smaller under both AR(1) ($85\%$) and AR(3)
($86\%$) than under AR(2) ($77\%$), rather than changing monotonically
with AR order.

\subsection{Type~I Error}
\label{sec:res-t1e}

\begin{figure}[htbp]
{\centering \includegraphics[height=0.8\textheight, width=\textwidth, keepaspectratio]{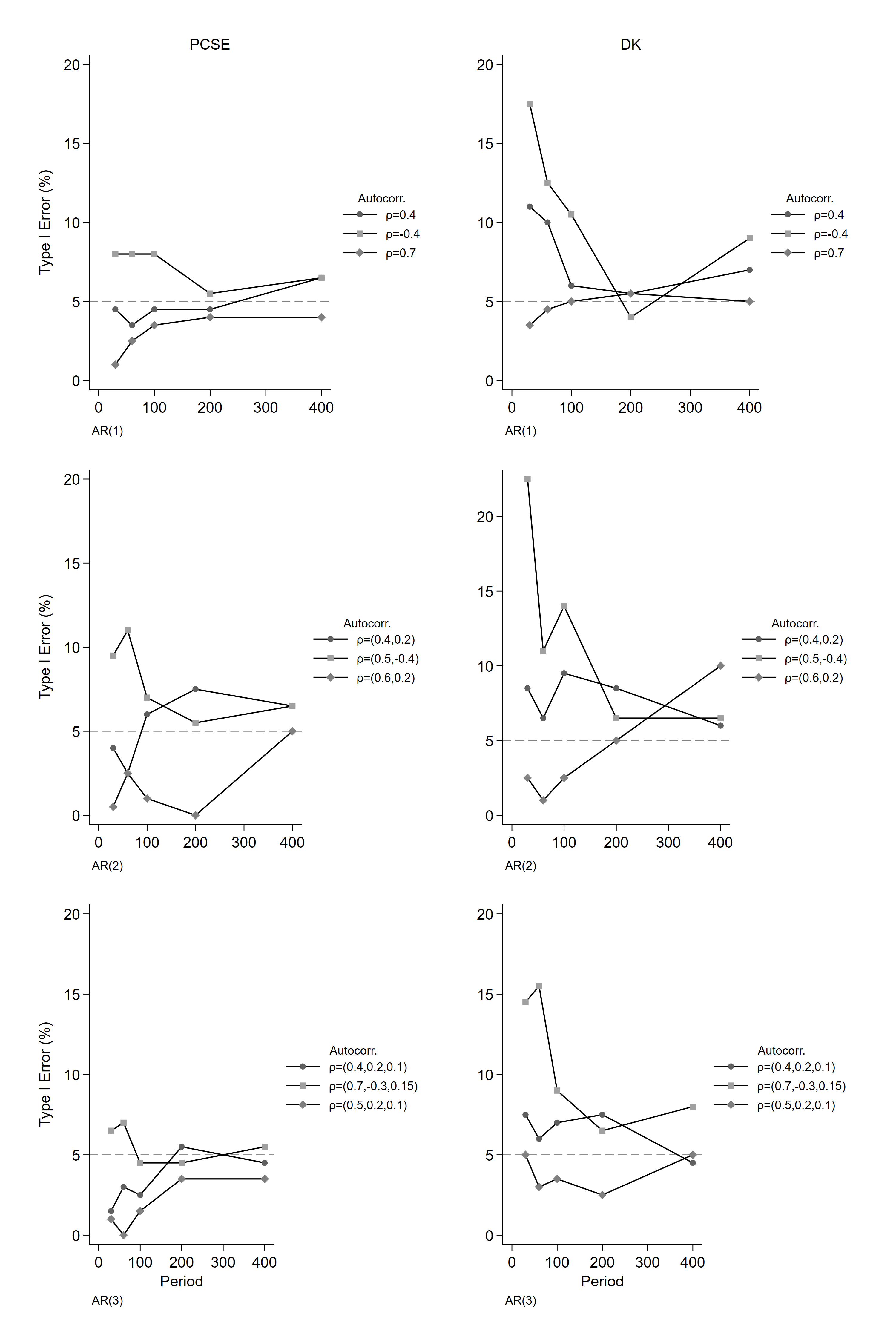}\par}
\caption{Type~I error for PCSE (left column) and DK (right column)
across all three autoregressive orders. Rows represent AR order.
Lines distinguish autocorrelation scenarios (AR-order-specific $\rho$
values as reported in Table~\ref{tab:simdesign}): mild positive
(circles); oscillatory (squares); high persistent (diamonds). Dashed
reference line at nominal $5\%$. $N=10$.}
\label{fig:3}
\end{figure}

All three autoregressive orders are presented together in
Figure~\ref{fig:3}, with rows representing AR order and columns
representing estimator. The naive (uncorrected) estimator was
reasonably well-calibrated under mild positive and high persistent
autocorrelation at every AR order (generally $0$--$6\%$), but showed
persistent, non-converging inflation under oscillatory
autocorrelation ($4$--$9\%$ regardless of $T$).

DK showed pronounced Type~I error inflation at short $T$ that grew
with autoregressive order: under oscillatory autocorrelation at
$T=30$, DK's Type~I error was $17.5\%$ at AR(1), $22.5\%$ at AR(2),
and $14.5\%$ at AR(3) -- elevated by a factor of three to four over
the nominal $5\%$ at every order. PCSE showed the same qualitative
pattern but consistently less severely -- $8$--$10\%$ at $T=30$
across AR orders under oscillatory autocorrelation, roughly half of
DK's inflation at the same cells -- and both corrections' Type~I
error moved toward nominal, without doing so smoothly at every
intermediate $T$, as $T$ increased. By $T=200$, both corrections were
within a few points of nominal across every scenario and AR order.

\subsection{Percentage Bias and RMSE}
\label{sec:res-bias}

Because PCSE and DK share identical point estimates, both bias and
RMSE are, by construction, identical between the two estimators; both
are therefore reported here numerically rather than graphically.
Under AR(2) error structures at $N=10$, percentage bias in the
treatment-effect coefficient was generally small and centered near
zero by $T \geq 100$ across all three scenarios, with no consistent
directional bias evident under oscillatory or mild positive
autocorrelation at any $T$. At short $T$ and the smallest effect size
($\Delta = 2\%$), percentage bias was occasionally large in magnitude
under high persistent autocorrelation (as large as $24\%$). RMSE
declined with $T$ across every
scenario and effect size, from roughly $0.001$--$0.003$ at $T=30$ to
under $0.0002$ by $T=400$ (logit scale), and did not show the
AR-order-dependent divergence seen in Type~I error and SE ratio.
AR(1) and AR(3) results followed the same pattern for both measures.

\subsection{Standard Error Ratio}
\label{sec:res-seratio}

\begin{figure}[htbp]
{\centering \includegraphics[height=0.8\textheight, width=\textwidth, keepaspectratio]{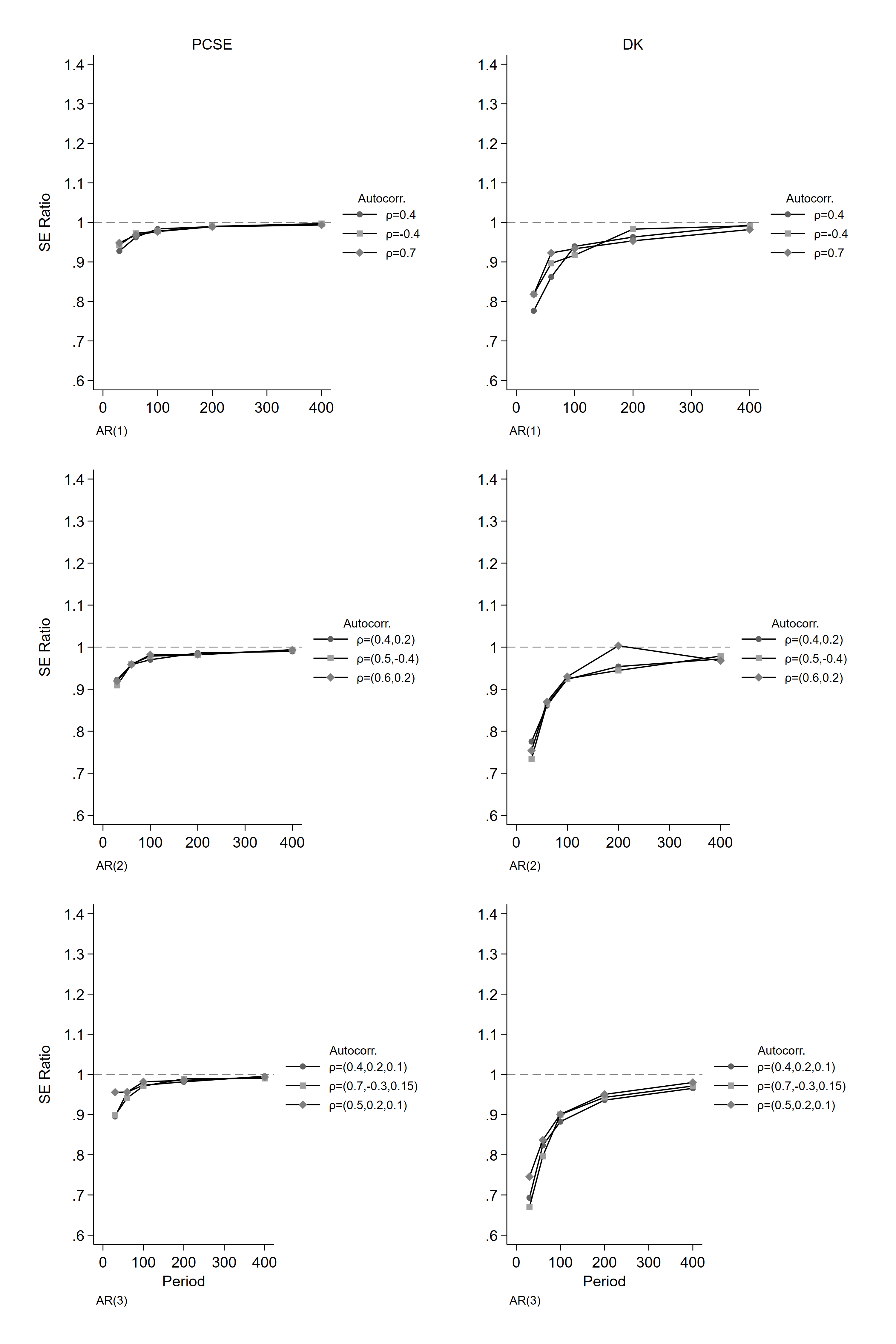}\par}
\caption{Standard error (SE) ratio for PCSE (left column) and DK
(right column) across all three autoregressive orders. Rows represent
AR order. Lines distinguish autocorrelation scenarios (AR-order-specific
$\rho$ values as reported in Table~\ref{tab:simdesign}): mild positive
(circles); oscillatory (squares); high persistent (diamonds). Dashed
reference line at $1.0$ (naive, exact by construction). Values below
$1.0$ indicate SE underestimation. $N=10$.}
\label{fig:6}
\end{figure}

As with Type~I error, all three autoregressive orders are presented
together in Figure~\ref{fig:6}. Naive's SE ratio is exactly $1.0$ in
every cell by construction, shown as a flat reference line rather
than a plotted trajectory. Both corrections underestimated the naive
standard error at short $T$, with the degree of underestimation
growing with autoregressive order: at $T=30$ under mild positive
autocorrelation, DK's SE ratio was $0.78$ at AR(1), $0.78$ at AR(2),
and $0.69$ at AR(3), while PCSE's was $0.93$, $0.92$, and $0.90$ over
the same orders. Both estimators' SE ratio moved toward $1.0$ as $T$
increased, with
PCSE closer to $1.0$ than DK at every AR order and $T$ examined: by
$T=400$, PCSE's SE ratio exceeded $0.99$ across all three AR orders,
while DK remained as low as $0.97$--$0.98$ under AR(3).

\subsection{Effect of Panel Size}
\label{sec:res-panelsize}

Supplement Figures~A5 and~A6 present Type~I error and SE
ratio under AR(2) error structures for $N \in \{10,15,20\}$, holding
$T$ on the horizontal axis as in every other figure in this section
and distinguishing panel size by line rather than by row. For both PCSE
and DK, the three panel-size curves tracked closely together across
every scenario and series length in both figures, with no consistent
trend toward better or worse calibration as $N$ increased from $10$
to $20$: the short-$T$ inflation and underestimation documented in
Sections~\ref{sec:res-t1e} and~\ref{sec:res-seratio}, and its
movement toward nominal as $T$ increased, appeared essentially the
same across all three panel sizes examined.

\subsection{Sensitivity Analysis}
\label{sec:res-sensitivity}

\begin{figure}[htbp]
\centering
{\centering \includegraphics[height=0.8\textheight, width=\textwidth, keepaspectratio]{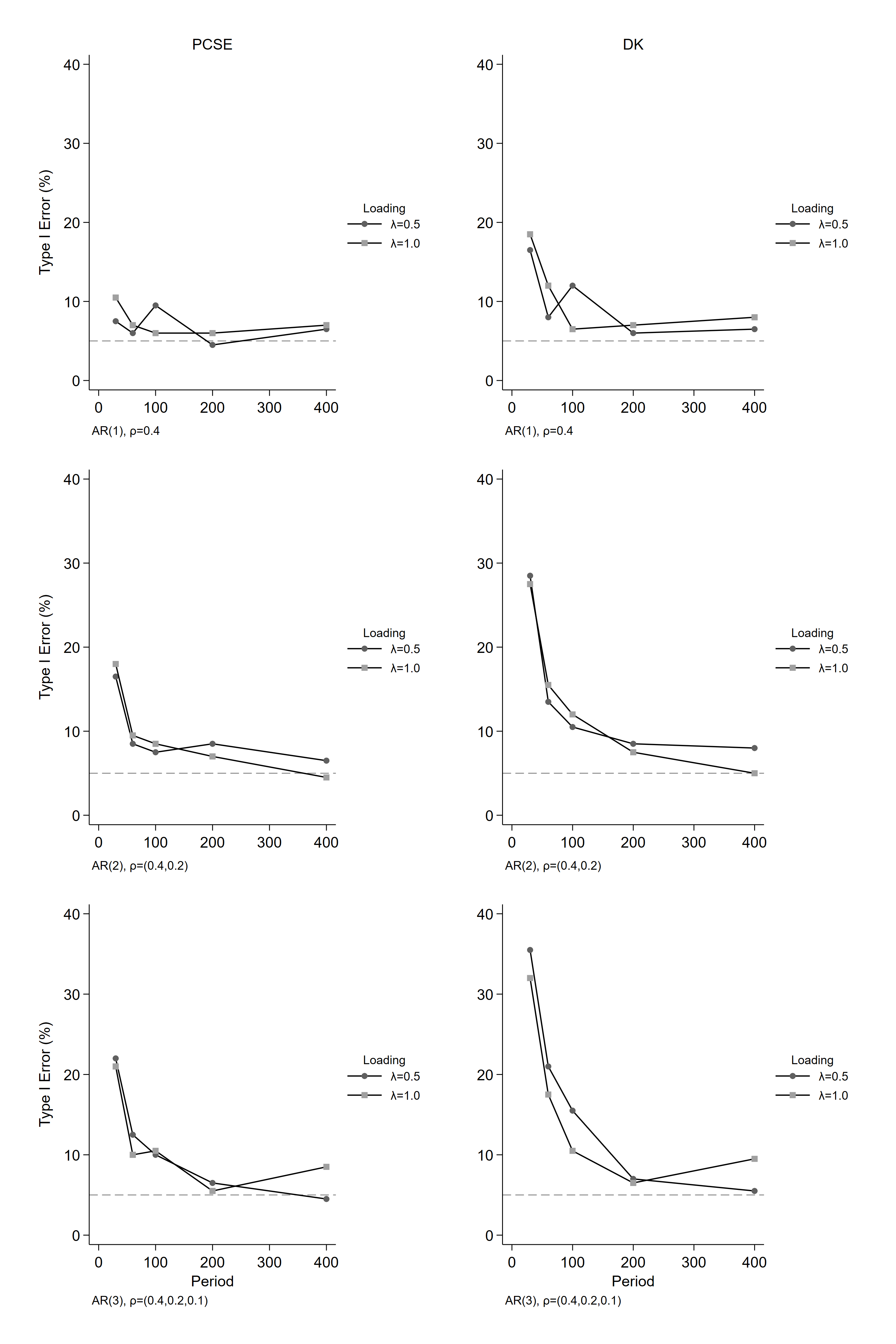}\par}
\caption{Type~I error for PCSE (left column) and DK (right column)
across all three autoregressive orders, under genuine cross-panel
dependence (mild positive autocorrelation scenario). Rows represent
AR order. Lines distinguish common-factor loading: $\lambda=0.5$
(circles); $\lambda=1.0$ (squares). Dashed reference line at nominal
$5\%$. Naive (uncorrected) is not shown; its Type~I error was
elevated and flat across all conditions (Section~\ref{sec:res-sensitivity}).
$N=10$.}
\label{fig:sens1}
\end{figure}

\begin{figure}[htbp]
\centering
{\centering \includegraphics[height=0.8\textheight, width=\textwidth, keepaspectratio]{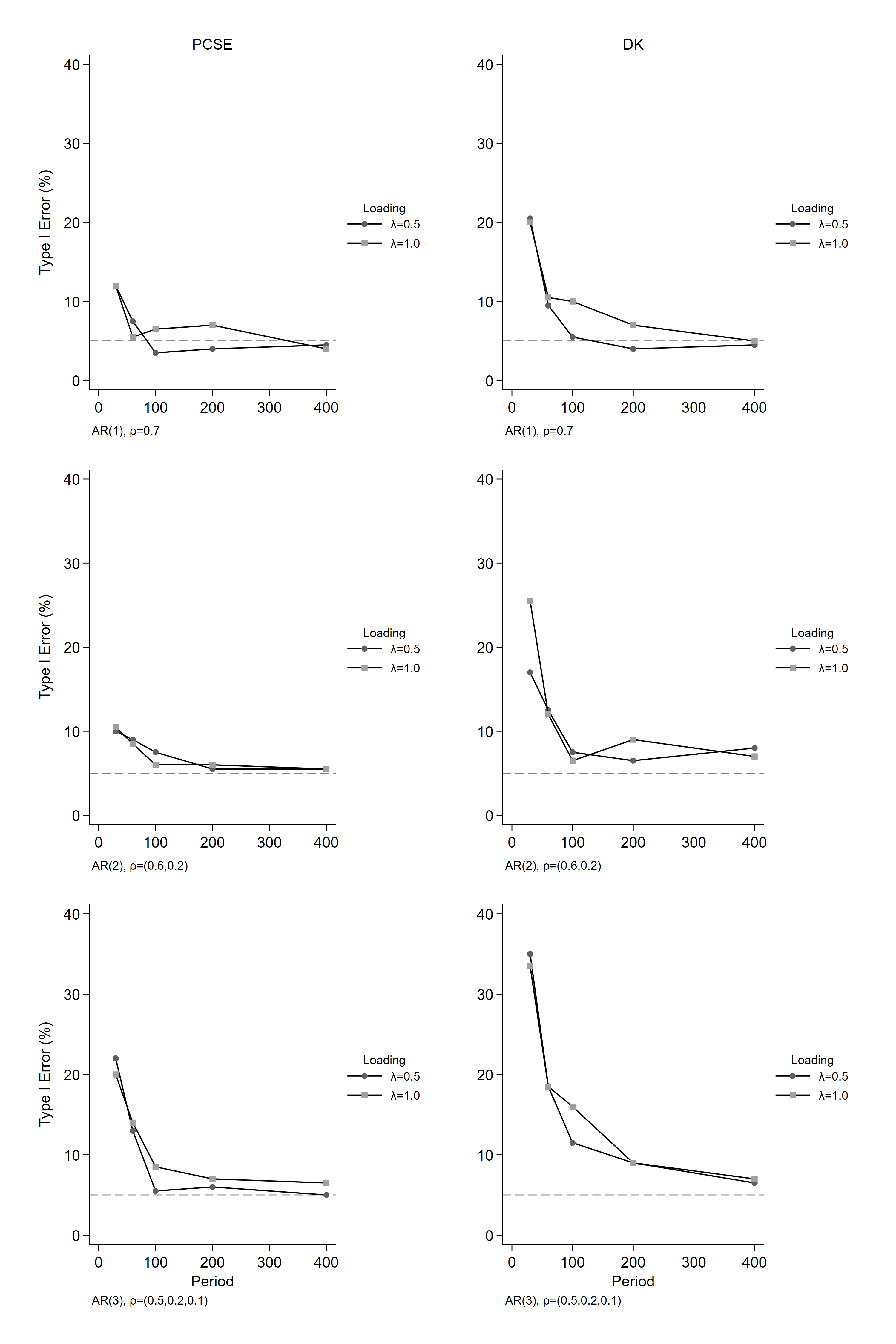}\par}
\caption{Type~I error for PCSE (left column) and DK (right column)
across all three autoregressive orders, under genuine cross-panel
dependence (high persistent autocorrelation scenario). Rows represent
AR order. Lines distinguish common-factor loading: $\lambda=0.5$
(circles); $\lambda=1.0$ (squares). Dashed reference line at nominal
$5\%$. Naive (uncorrected) is not shown; its Type~I error was
elevated and flat across all conditions (Section~\ref{sec:res-sensitivity}).
$N=10$.}
\label{fig:sens2}
\end{figure}

\begin{figure}[htbp]
\centering
{\centering \includegraphics[height=0.8\textheight, width=\textwidth, keepaspectratio]{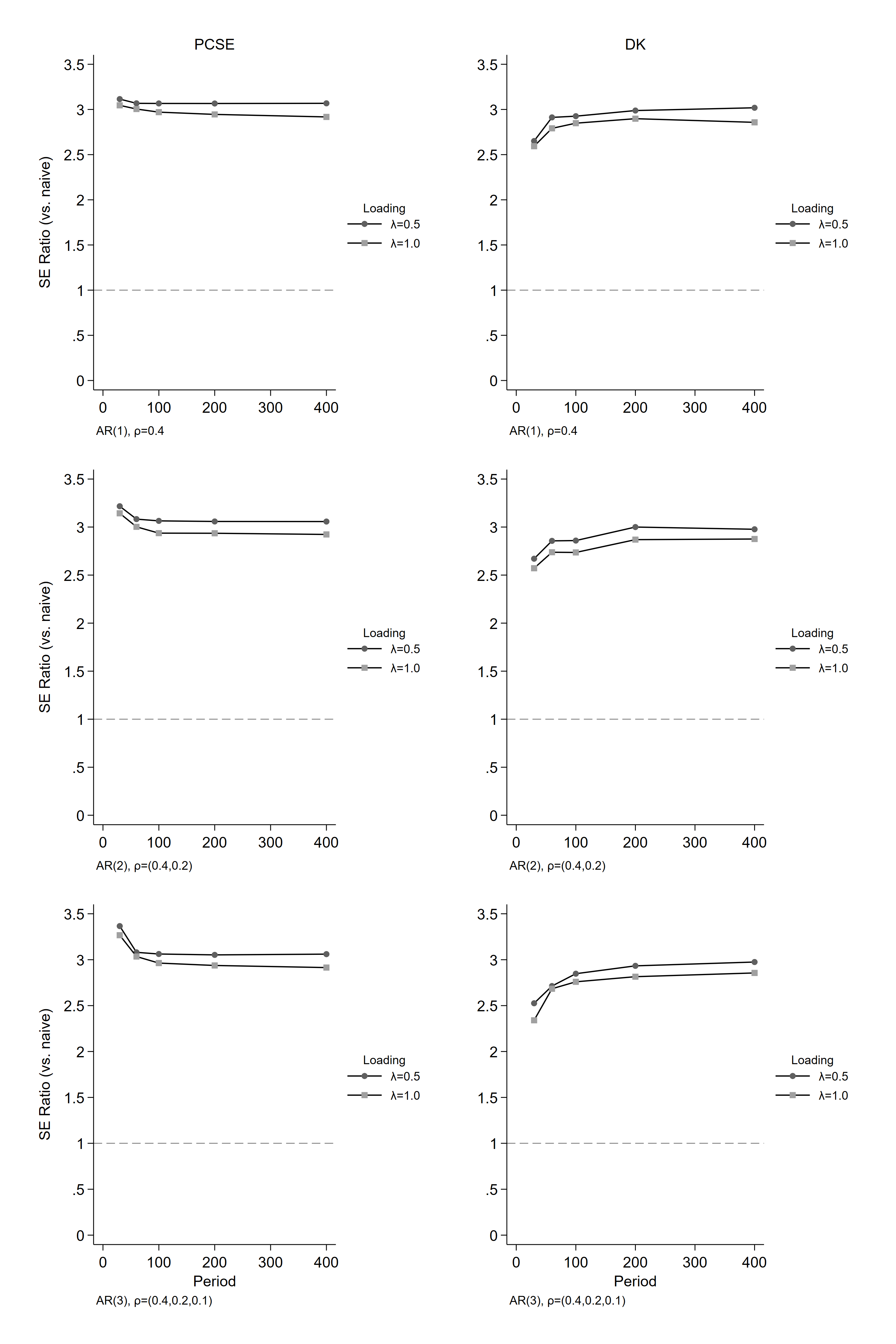}\par}
\caption{Standard error (SE) ratio versus naive for PCSE (left
column) and DK (right column) across all three autoregressive
orders, under genuine cross-panel dependence (mild positive
autocorrelation scenario). Rows represent AR order. Lines distinguish
common-factor loading: $\lambda=0.5$ (circles); $\lambda=1.0$
(squares). Dashed reference line at $1.0$: unlike the primary design,
values here are well above $1.0$ throughout, reflecting naive's own
underestimated standard error under genuine cross-panel dependence,
not underestimation by either correction. $N=10$.}
\label{fig:sens3}
\end{figure}

\begin{figure}[htbp]
\centering
{\centering \includegraphics[height=0.8\textheight, width=\textwidth, keepaspectratio]{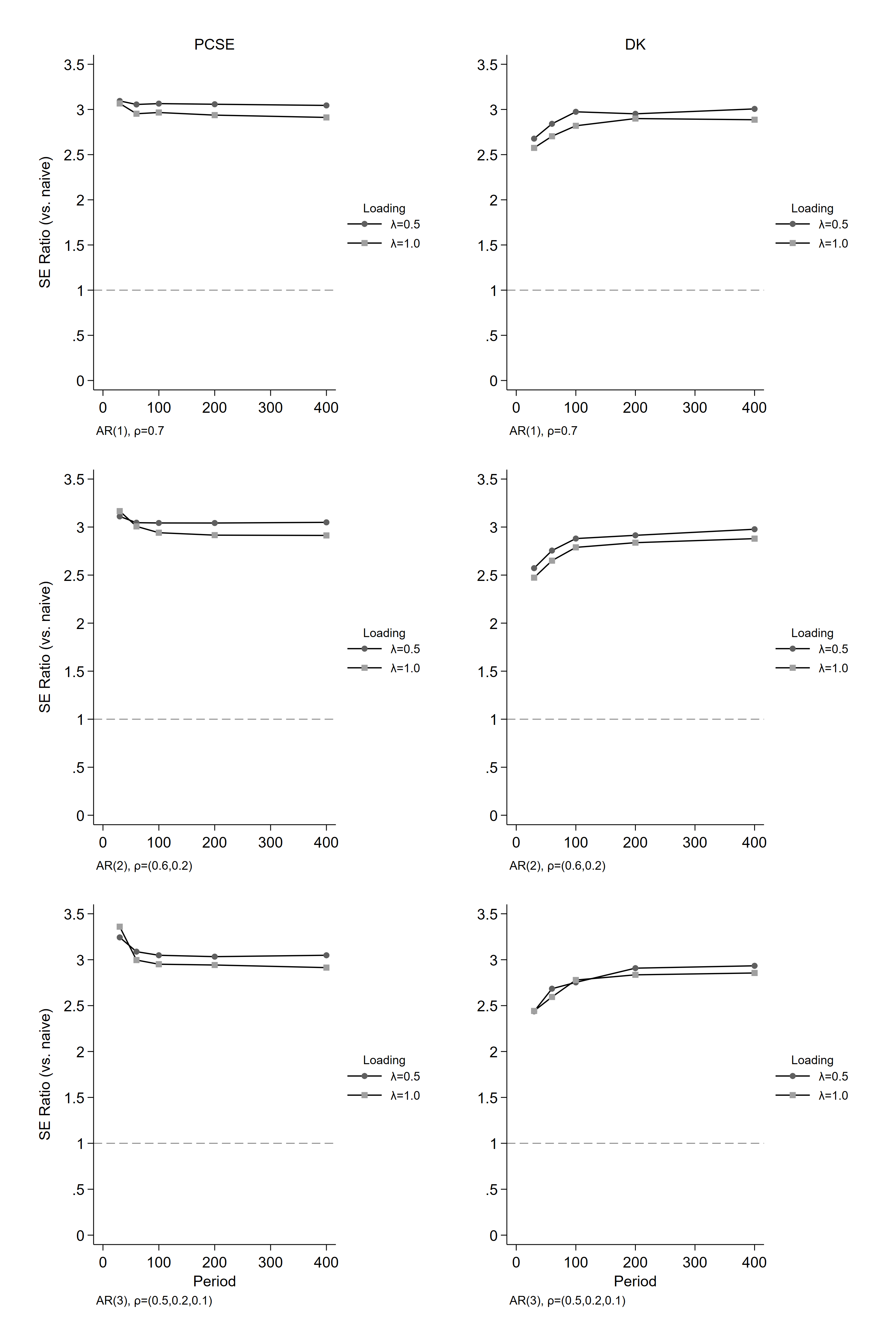}\par}
\caption{Standard error (SE) ratio versus naive for PCSE (left
column) and DK (right column) across all three autoregressive
orders, under genuine cross-panel dependence (high persistent
autocorrelation scenario). Rows represent AR order. Lines distinguish
common-factor loading: $\lambda=0.5$ (circles); $\lambda=1.0$
(squares). Dashed reference line at $1.0$: unlike the primary design,
values here are well above $1.0$ throughout, reflecting naive's own
underestimated standard error under genuine cross-panel dependence,
not underestimation by either correction. $N=10$.}
\label{fig:sens4}
\end{figure}

Figures~\ref{fig:sens1}--\ref{fig:sens4} present Type~I error and SE ratio
under genuine cross-panel dependence, introduced via a common factor
with loading $\lambda \in \{0.5,1.0\}$ (Section~\ref{sec:simstrategy}).
Naive's Type~I error is not shown graphically: it was elevated and
essentially flat across every AR order, scenario, loading, and series
length examined, ranging from $45.5\%$ to $76.5\%$ (nominal $5\%$),
with no tendency to approach nominal as $T$ increased from $30$ to
$400$.

Both PCSE and DK showed Type~I error elevated at short $T$ that moved
toward nominal as $T$ increased, at every AR order and under both
scenarios. Under high persistent autocorrelation at $N=10$, DK's
Type~I error fell from $20$--$34\%$ at $T=30$ to $5$--$8\%$ at
$T=400$ across the three AR orders; PCSE's fell from $10$--$22\%$ to
$4$--$6\%$ over the same range, consistently lower than DK's at every
AR order and $T$ examined. The same pattern held under mild positive
autocorrelation.

SE ratio was substantially above $1.0$ throughout -- $2.3$ to $4.6$
across all conditions -- reflecting the naive estimator's own
standard error rather than either correction's. At $N=10$, high
persistent autocorrelation, AR(2), DK's SE ratio was $2.52$ at $T=30$
and $2.93$ at $T=400$; PCSE's was $3.14$ at $T=30$ and $2.98$ at
$T=400$, converging to a similar magnitude by $T=400$ despite PCSE
starting from a larger ratio at $T=30$.

\section{Illustrative Example}
\label{sec:example}

To illustrate the practical consequences of estimator choice in a
realistic panel setting, we extend \texttt{betark}'s own single-practice
illustrative example \cite{lindenbetarkpaper2026} to a multi-site
disease management program. Disease management programs target
individuals with chronic conditions through structured behavioral and
clinical interventions \cite{linden2008,linden2003dmaa,lindenblackbox2006},
and prediabetes is a common target given the effectiveness of
lifestyle interventions in preventing progression to type~2 diabetes
\cite{biuso2007}.

We do not have access to real multi-site panel data of this kind;
the example below uses simulated data solely to demonstrate the use
of the software, not to showcase the estimator resolving a genuinely
messy empirical problem. The artificial study involves $N=12$
primary care practices whose
enrolled prediabetic patients were fitted with continuous glucose
monitors and tracked daily for $T=300$ days. The outcome is the daily
proportion of each practice's enrolled patients whose fasting glucose
meets or exceeds the clinical target of $100$ mg/dL, aggregated across
each practice's patient panel. At day $150$, all twelve practices
introduced the same comprehensive lifestyle intervention program
simultaneously. Panels share an identical deterministic mean
structure: a starting proportion of $\mu_0=.35$, a $10\%$ relative
increase in the elevated-glucose proportion over the pre-intervention
period, no immediate level change at the intervention, and a $15\%$
relative decrease over the post-intervention period -- the same
parameterization \texttt{betark}'s own example used, deliberately
chosen there to represent a modest, clinically plausible improvement
rather than an effect large enough to remain significant regardless
of standard error calibration. Dispersion was set to $cv=.05$,
matching the primary simulation. Panels differ only in their AR($k$)
innovations.

High persistent positive autocorrelation was specified to reflect
plausible serial dependence in longitudinal, practice-level aggregate
glycemic measurements. A single realization (not an average over
replications) is generated under each of three autoregressive orders,
using a common seed and the same high-persistent scenario values as
the primary simulation: $\rho=0.7$ at AR(1), $\boldsymbol{\rho}=(0.6,0.2)$
at AR(2), and $\boldsymbol{\rho}=(0.5,0.2,0.1)$ at AR(3). These differ
from the high-persistent values used in \texttt{betark}'s own
single-series example ($\rho=(0.7,0.2)$ and $(0.6,0.25,0.1)$
respectively); we use our own revised values throughout, for the same
reason given in Section~\ref{sec:res-t1e} -- the original values sit
close enough to the AR($k$) stationarity boundary that an individual
realization can occasionally produce a degenerate fit.

\texttt{xtbetark} was fit to each realization with \texttt{lag()} set
to the true AR order, once under each variance estimator
(\texttt{vce(oim)}, \texttt{vce(dk)}, and \texttt{vce(pcse)}).
Because the data-generating process is identical across methods for a
given AR order, the fitted trajectories do not differ between them;
the methods diverge only in their standard errors. Table~\ref{tab:applied}
reports the resulting slope-change coefficient and its three standard
errors. The coefficient is, as expected, essentially identical across
methods within each AR order. The three standard errors are not: DK's
confidence interval is narrower than both naive's and PCSE's at every
AR order, most visibly at AR(3), where DK's interval width is roughly
$17\%$ smaller than
PCSE's. This single realization reproduces, concretely, the systematic
pattern documented across the full simulation in Section~\ref{sec:results}.

\section{Discussion}
\label{sec:discussion}

The fundamental finding of this study is that PCSE and DK do not
simply trade calibration for power against one another: PCSE achieved
better-calibrated inference than DK at every autoregressive order
examined, at a power cost that was modest and, where DK held a
nominal edge, at least partly illusory given DK's own simultaneous
miscalibration. This mirrors the relationship
documented between PCSE and the
Driscoll--Kraay estimator for continuous outcomes \cite{lindenxtpraiskpaper2026}, and extends it to
a setting neither that work nor the single-series beta-AR($k$) model
\cite{lindenbetarkpaper2026} addressed alone: a bounded,
nonlinearly-estimated outcome pooled across panels. Table~\ref{tab:summary}
summarizes performance across AR orders.

\subsection{Statistical Power}

Power was highest under AR(1), with AR(2) and AR(3) closer to each
other than either was to AR(1) and no consistent ordering between
them (Section~\ref{sec:res-power}) -- a more complicated picture than
the clean, progressively widening AR-order gradient
reported for the analogous comparison
between PCSE and the Driscoll--Kraay estimator for continuous outcomes \cite{lindenxtpraiskpaper2026}. A plausible
explanation is estimation method rather than the panel extension
itself: unlike the closed-form Prais--Winsten GLS transformation
compared against the Driscoll--Kraay
estimator there \cite{lindenxtpraiskpaper2026}, the beta-AR($k$) likelihood here is fit by iterative
Newton--Raphson optimization, and a bounded outcome's values landing
close to $0$ or $1$ can affect a given replication's convergence in a
way a linear estimator's direct algebraic solution cannot. An
alternative hypothesis is that our five-point $T$ grid, coarser than
the six-point grid used previously \cite{lindenxtpraiskpaper2026}, makes ordinary
sampling noise visually resemble discontinuities a finer grid would
smooth over; distinguishing between these two hypotheses would
require a matched, finer-grained $T$ sequence. DK's
short-$T$ power edge over PCSE should be read against
Section~\ref{sec:disc-coverage} below: a method that rejects the null
too often will also appear to reject more often under a true
alternative, without that reflecting a more sensitive test.

\subsection{Coverage and Type~I Error}
\label{sec:disc-coverage}

DK's inflation and coverage deficits were concentrated at short $T$
and diminished as $T$ grew (Section~\ref{sec:res-t1e}), the same
finite-sample-concentrated pattern
reported for the Driscoll--Kraay estimator previously \cite{lindenxtpraiskpaper2026}. This is consistent with
the case developed in Section~\ref{sec:dk}: because the beta-AR($k$)
likelihood already removes serial dependence parametrically, DK's
kernel may be smoothing over structure already accounted for, so its
miscalibration would be expected to behave like ordinary finite-sample noise rather
than a persistent problem -- which is what both this study and that
earlier comparison independently observed \cite{lindenxtpraiskpaper2026}. The
simulation results are consistent with this mechanism rather than a
direct test of it.

Unlike Power, this pair of measures did show AR(2) as a specific
outlier rather than the three AR orders forming a mixed pattern
throughout -- a departure from the
progressively worsening gradient reported previously \cite{lindenxtpraiskpaper2026}. We hypothesize this may reflect the
same estimation-method distinction discussed above, though this
remains to be tested directly. Naive's persistent,
non-converging Type~I error inflation under oscillatory
autocorrelation specifically (Section~\ref{sec:res-t1e}) is
consistent with a structural rather than finite-sample problem:
naive's covariance has no mechanism to absorb the correlation the
AR($k$) likelihood itself is designed to model.

\subsection{Bias, RMSE, and SE Ratio}

PCSE and DK share identical point estimates by construction, so bias
and RMSE cannot differ between them; both were essentially unbiased
by $T \geq 100$, paralleling an
analogous finding for continuous outcomes \cite{lindenxtpraiskpaper2026} and suggesting neither the panel extension nor its
adaptation to a bounded outcome introduces material bias. RMSE was
the one measure that did follow a clean AR-order gradient
(Section~\ref{sec:res-bias}), consistent with each additional
autoregressive term adding estimation burden to the joint likelihood.

SE ratio is consistent with why DK's
Type~I error and coverage were more affected than PCSE's: DK's
underestimation of naive's standard error widened with AR order while
PCSE's did not (Section~\ref{sec:res-seratio}) -- the pattern
motivating DK's inclusion as a comparator in
Section~\ref{sec:dk}, and consistent with
an analogous finding that the
Driscoll--Kraay estimator's SE underestimation worsened with AR order
while PCSE's did not \cite{lindenxtpraiskpaper2026}.

\subsection{Effect of Panel Size}

Type~I error and SE ratio were essentially unaffected by panel size
across $N=10$ to $N=20$ (Section~\ref{sec:res-panelsize}) -- the same
conclusion reached previously for PCSE and the
Driscoll--Kraay estimator \cite{lindenxtpraiskpaper2026}. Replicating this in a structurally
different model with a bounded rather than continuous outcome
suggests that series length, not panel count,
governs panel-corrected inference under serial dependence generally,
rather than being specific to a linear GLS framework.

\subsection{Cross-Panel Dependence}
\label{sec:disc-crosspanel}

The sensitivity analysis evaluating genuine cross-panel dependence
confirms the central premise motivating this paper: naive's Type~I
error was not merely elevated but structurally broken, showing no
tendency toward nominal performance even at $T=400$
(Section~\ref{sec:res-sensitivity}), unlike every miscalibration
documented under independent panels, all of which shrank with $T$.
Both PCSE and DK corrected this, with PCSE again better calibrated
than DK at every AR order and series length -- extending, to the one
condition where a panel correction is not merely helpful but
necessary, the same advantage documented throughout the independent-panel
results above. This mirrors a previous finding that PCSE remained robust to a common-factor loading \cite{lindenxtpraiskpaper2026}.

One pattern in the SE ratio results of the sensitivity analysis 
differs from that of the primary analysis in which
PCSE's ratio versus naive was occasionally higher than DK's at short
$T$ before the two converged to a similar magnitude by $T=400$, the
opposite of PCSE's usual advantage in absolute calibration. We
hypothesize this may reflect a difference in how the two corrections
estimate cross-panel covariance when little information is available:
PCSE's contemporaneous sample covariance may be more variable with
few periods to estimate it from, plausibly overstating the correction
needed at short $T$, while DK's kernel-smoothed estimate is more
conservative until enough periods accumulate for its bandwidth to
capture the full correlation structure.

\subsection{Practical Recommendations}

PCSE is the preferred variance estimator for \texttt{xtbetark} across
the conditions examined: better-calibrated Type~I error and coverage
than DK at every AR order and series length, at a power cost that was
modest and partly illusory at short $T$. This advantage was most
pronounced at short $T$ and high AR order -- conditions common in
panel-ITS applications with monthly or weekly observation. Researchers
working with $T < 60$ under suspected higher-order autocorrelation
should be particularly cautious about DK specifically. The naive
estimator's persistent oscillatory-scenario Type~I error problem,
regardless of series length, makes it a risky default even when
cross-panel independence seems plausible. Researchers should assess the autocorrelation structure of their data, report the lag order used, and consider sensitivity analyses under alternative specifications \cite{linden2005}.

\subsection{Limitations}
\label{sec:disc-limitations}

Several limitations of the present study should be noted. The
simulations were restricted to balanced panels, a single covariate,
shared intervention timing, and independent AR($k$) innovations in
the primary design; these restrictions may not reflect the full
range of applied panel data structures. The factor model used in the
sensitivity analysis represents one approach to cross-panel
dependence; other forms of spatial or network dependence were not
examined. Effect sizes were confined to $\{2\%,4\%,8\%\}$, calibrated
to avoid the power saturation observed at larger magnitudes in
preliminary testing, leaving larger effects unexamined. The maximum
AR order examined was $3$; whether PCSE's calibration advantage and
DK's small-$T$ inflation persist at AR(4) and beyond is unknown.

Only two panel-corrected variance estimators were compared. This
choice was intentional, as the objective of the study was to compare
parametric and nonparametric approaches to handling serial dependence
within the same joint-likelihood inferential framework, analogous to
the comparison for continuous outcomes \cite{lindenxtpraiskpaper2026}.
Other modeling strategies for proportion and rate outcomes exist and
were beyond the scope of this study: machine learning approaches
have been applied to related interrupted time series problems
\cite{lindenyarnold2018}; panel fractional response models estimate
proportion outcomes via pooled quasi-maximum likelihood with
cluster-robust standard errors rather than a joint likelihood
\cite{papke2008}; and Gaussian copula-based beta regression models
serial dependence in bounded time series through the correlation
structure directly rather than a recursive conditional likelihood
\cite{guolovarin2014}. None of these addresses the specific
combination of a beta-distributed panel outcome with AR($k$) errors
that motivates this paper, but each represents a methodological
direction future comparative work could pursue. DK's bandwidth was
likewise fixed a priori at $L=k$ on theoretical grounds
(Section~\ref{sec:dk}) rather than compared against alternative
bandwidths within this study; a direct sensitivity comparison is a
natural next step and would either strengthen or narrow the case made
here for DK's short-$T$ anticonservative bias.

\section{Conclusion}
\label{sec:conclusion}

To our knowledge, no existing likelihood-based estimator jointly
models beta-distributed panel outcomes with arbitrary-order
autoregressive errors while providing panel-specific covariance
corrections. This paper
developed and validated one: a joint conditional maximum-likelihood
approach for beta-distributed panel outcomes with AR($k$) errors,
paired with a choice between two panel-appropriate variance
estimators. Across three autoregressive orders and both independent
panels and genuine cross-panel dependence, the parametric variance
estimator (PCSE) was consistently better calibrated than its
nonparametric alternative (DK), at a power cost that was modest and
partly illusory. The advantage was largest precisely where it
matters most: under genuine cross-panel dependence, an uncorrected
estimator failed in a way that additional data could not fix, while
both corrections resolved it.

For applied researchers working with bounded panel outcomes and
suspected serial or cross-panel dependence -- common in multi-site
disease management, quality-improvement, and public health
surveillance data -- this work provides both the methodology and a
validated basis for choosing between the two corrections available
to implement it, with PCSE the recommended default. The limitations
noted above, particularly the comparison to other modeling
strategies for proportion and rate outcomes, mark the most promising
directions for extending this work.

\bibliographystyle{tfnlm}
\bibliography{xtbetark_arxiv_refs}

\clearpage

\section*{Abbreviations}

{\sloppy
AR: autoregressive; BHHH: Berndt-Hall-Hall-Hausman; CI: confidence
interval; CSTS: cross-sectional time-series; DK: Driscoll-Kraay;
GLM: generalized linear model; GLS: generalized least squares; HAC:
heteroskedasticity- and autocorrelation-consistent; ITS: interrupted
time series; MLE: maximum likelihood estimator; OIM: observed
information matrix; OLS: ordinary least squares; PCSE:
panel-corrected standard errors; RMSE: root mean squared error; SE:
standard error.
\par}

\bigskip
\paragraph{Supplemental data.} Supplemental data for this article can be accessed online at XXXXXXXX.

{\sloppy
\paragraph{Data availability.} Stata code used in this paper is found at: \url{https://github.com/ariellinden/xtbetark}
\par}

\clearpage

\begin{table}[htbp]
\centering
\caption{Simulation design parameters for the primary and sensitivity analyses}
\label{tab:simdesign}
\begin{tabular}{ll}
\toprule
Design parameter & Values \\
\midrule
Autoregressive order, $k$ & 1, 2, 3 \\
Autocorrelation scenario & Mild positive, Oscillatory\textsuperscript{a}, High persistent \\
\quad AR(1) coefficient & $0.4$, $-0.4$, $0.7$ \\
\quad AR(2) coefficients & $(0.4,0.2)$, $(0.5,-0.4)$, $(0.6,0.2)$ \\
\quad AR(3) coefficients & $(0.4,0.2,0.1)$, $(0.7,-0.3,0.15)$, $(0.5,0.2,0.1)$ \\
Number of panels, $N$ & 10, 15, 20 \\
Series length, $T$ & 30, 60, 100, 200, 400 \\
Post-intervention effect size & 0\%, 2\%, 4\%, 8\% \\
Common factor loading, $\lambda$\textsuperscript{b} & 0.5, 1.0 \\
Starting mean, $\mu_0$ & 0.10 \\
Coefficient of variation & 0.05 \\
Pre-intervention trend & 0\% \\
Immediate post-intervention level change & 0\% \\
Replications per design cell & 2{,}000 \\
\bottomrule
\end{tabular}

\vspace{2mm}
\begin{minipage}{0.95\textwidth}
\footnotesize
\textsuperscript{a}Oscillatory scenario used only in the primary
design; the sensitivity design uses the mild positive and
high-persistent scenarios only. \textsuperscript{b}Applies only to the
sensitivity design (genuine cross-panel dependence); the primary
design uses independent panels ($\lambda = 0$, not shown).
\end{minipage}
\end{table}

\clearpage

\renewcommand{\arraystretch}{1.3}
\begin{table}[htbp]
\centering
\footnotesize
\caption{Summary of simulation findings across AR(1), AR(2), and AR(3) error structures}
\label{tab:summary}
\begin{tabular}{p{1.9cm}p{4.15cm}p{4.15cm}p{4.15cm}}
\toprule
Measure & AR(1) & AR(2) & AR(3) \\
\midrule

Power &
DK modestly higher than PCSE at short $T$ (e.g., $T=30$, $\Delta=8\%$,
high persistent: DK $57\%$, PCSE $48\%$). Highest power of the three
AR orders for a given $T$ and effect size. &
DK modestly higher than PCSE at short $T$ ($T=30$, $\Delta=8\%$, high
persistent: DK $46\%$, PCSE $26\%$). Lowest power of the three AR
orders at several $T$/effect-size combinations. &
DK modestly higher than PCSE at short $T$ ($T=30$, $\Delta=8\%$, high
persistent: DK $55\%$, PCSE $32\%$). Similar to AR(2); no consistent
ordering between AR(2) and AR(3). \\

95\% Coverage &
Near nominal for mild positive and high persistent scenarios.
Oscillatory departure at $T=30$: DK $85\%$, PCSE $92\%$. &
Near nominal for mild positive and high persistent scenarios.
Oscillatory departure at $T=30$ largest of the three AR orders: DK
$77\%$, PCSE $85$--$91\%$. &
Near nominal for mild positive and high persistent scenarios.
Oscillatory departure at $T=30$: DK $86\%$, PCSE $95\%$ -- similar to
AR(1), not larger than AR(2). \\

Type~I Error &
Naive $0$--$7\%$ across scenarios. Oscillatory inflation at $T=30$:
DK $17.5\%$, PCSE $8.0\%$. &
Naive $0$--$7\%$ across scenarios. Oscillatory inflation at $T=30$
largest of the three AR orders: DK $22.5\%$, PCSE $9.5\%$. &
Naive $0$--$4\%$ across scenarios. Oscillatory inflation at $T=30$:
DK $14.5\%$, PCSE $6.5\%$ -- similar to AR(1), not larger than
AR(2). \\

Bias &
Generally small by $T \geq 100$; maximum magnitude at $\Delta=2\%$
across all scenarios and $T$: $12\%$. &
Generally small by $T \geq 100$; maximum magnitude at $\Delta=2\%$:
$24\%$, the largest of the three AR orders. &
Generally small by $T \geq 100$; maximum magnitude at $\Delta=2\%$:
$21\%$ -- similar to AR(2). \\

RMSE &
Declines with $T$ (e.g., $\Delta=4\%$: $0.0014$ at $T=30$ to
$<0.0001$ by $T=400$, logit scale). Lowest of the three AR orders at
a given $T$. &
Declines with $T$ ($\Delta=4\%$: $0.0019$ at $T=30$ to $0.0001$ by
$T=400$). Intermediate between AR(1) and AR(3). &
Declines with $T$ ($\Delta=4\%$: $0.0021$ at $T=30$ to $0.0001$ by
$T=400$). Highest of the three AR orders at a given $T$. \\

SE Ratio &
DK underestimates naive's SE more than PCSE at short $T$ (mild
positive, $T=30$: DK $0.78$, PCSE $0.93$). Both approach $1.0$ by
$T=400$ ($\geq 0.99$). &
DK underestimates naive's SE more than PCSE at short $T$ (mild
positive, $T=30$: DK $0.78$, PCSE $0.92$). Both approach $1.0$ by
$T=400$. &
DK underestimates naive's SE more than PCSE at short $T$, most of
the three AR orders (mild positive, $T=30$: DK $0.69$, PCSE $0.90$).
DK remains below $0.99$ at $T=400$. \\

\bottomrule
\end{tabular}

\vspace{2mm}
\begin{minipage}{0.95\textwidth}
\footnotesize
Cell entries summarize patterns described in full in
Section~\ref{sec:results}; all figures are for $N=10$. Power,
Coverage, and Type~I error entries for AR(1) and AR(3) are drawn from
the corresponding Supplement figures. Note that AR(2) is not
consistently intermediate between AR(1) and AR(3): for Power,
Coverage, Type~I error, and Bias, AR(2) or the AR(2)/AR(3) pair
showed the largest departure from nominal performance among the
three orders examined, rather than departure increasing smoothly
with AR order. RMSE is the exception, increasing monotonically with
AR order at every $T$ examined.
\end{minipage}
\end{table}
\clearpage

\begin{table}[htbp]
\centering
\caption{Illustrative example: slope-change coefficient by AR order and estimation method}
\label{tab:applied}
\begin{tabular}{llrrrr}
\toprule
AR Order & Method & Coefficient & SE & $z$ & 95\% CI \\
\midrule
AR(1) & naive & $-0.00270$ & $0.000198$ & $-13.65$ & $(-0.00309,\ -0.00231)$ \\
      & DK    & $-0.00270$ & $0.000174$ & $-15.55$ & $(-0.00304,\ -0.00236)$ \\
      & PCSE  & $-0.00270$ & $0.000183$ & $-14.71$ & $(-0.00306,\ -0.00234)$ \\
\addlinespace
AR(2) & naive & $-0.00271$ & $0.000289$ & $-9.39$  & $(-0.00328,\ -0.00215)$ \\
      & DK    & $-0.00271$ & $0.000257$ & $-10.58$ & $(-0.00322,\ -0.00221)$ \\
      & PCSE  & $-0.00271$ & $0.000286$ & $-9.49$  & $(-0.00328,\ -0.00215)$ \\
\addlinespace
AR(3) & naive & $-0.00268$ & $0.000282$ & $-9.50$  & $(-0.00324,\ -0.00213)$ \\
      & DK    & $-0.00268$ & $0.000243$ & $-11.04$ & $(-0.00316,\ -0.00221)$ \\
      & PCSE  & $-0.00268$ & $0.000294$ & $-9.11$  & $(-0.00326,\ -0.00211)$ \\
\bottomrule
\end{tabular}

\vspace{2mm}
\begin{minipage}{0.95\textwidth}
\footnotesize
Coefficient is the estimated slope change (day$^{-1}$) following the
intervention, on the logit scale, for a single realization at $N=12$,
$T=300$. Identical across methods within each AR order by
construction (Section~\ref{sec:res-bias}); methods differ only in
their standard error.
\end{minipage}
\end{table}

\renewcommand{\arraystretch}{1}

\clearpage

\end{document}